\documentclass[11pt]{article}
\usepackage{amstext,amssymb,amsmath,amsbsy}

\usepackage{epstopdf}
\usepackage{epsfig}
\usepackage{hyperref}
\usepackage{amscd}
\usepackage{amsfonts}
\usepackage{indentfirst}
\usepackage{verbatim}
\usepackage{amsmath}
\usepackage{color}
\usepackage{amsthm}
\usepackage{enumerate}
\usepackage{graphicx}
\usepackage{subfig}
\usepackage[OT1]{fontenc}
\usepackage[latin1]{inputenc}
\usepackage[english]{babel}
\usepackage{amssymb}
\newtheorem{theorem}{Theorem}
\newtheorem{lemma}{Lemma}

\newtheorem{corollary}{Corollary}
\newtheorem{remark}{Remark}

\newtheorem{definition}{Definition}
\numberwithin{equation}{section}

\newcommand{\proofend}{\hfill $\Box$ }

\newcommand{\ess}{\mathrm{ess} \,}

\newcommand{\supp}{\operatorname{supp}}

\newcommand{\dive}{\operatorname{div}}

\newcommand{\eps}{\varepsilon}

\newcommand{\epss}{s}

\newcommand{\loc}{_{loc}}

\newcommand{\mR}{\mathbb{R}}

\title{Asymptotic behavior of solutions to the Helmholtz equations with sign changing coefficients}

\author{Hoai-Minh Nguyen \footnote{EPFL SB MATHAA CAMA, Station 8,  CH-1015 Lausanne, hoai-minh.nguyen@epfl.ch.} \footnote{School of Mathematics, University of Minnesota, Minneapolis, MN, 55455, hmnguyen@math.umn.edu.} \footnote{The research is supported by NSF grant DMS-1201370 and by the Alfred P. Sloan Foundation.}}

\begin{document}

\maketitle

\begin{abstract}
This paper is devoted to the study of the behavior of the unique solution $u_\delta \in H^{1}_{0}(\Omega)$,  as $\delta \to 0$,  to the equation
\begin{equation*}
\dive(\epss_\delta A \nabla u_{\delta}) + k^2 \epss_0 \Sigma u_{\delta} = \epss_0 f \mbox{ in } \Omega,
\end{equation*}
where $\Omega$ is a smooth connected bounded open subset of $\mR^d$  with $d=2$ or 3, $f \in L^2(\Omega)$, $k$ is a non-negative constant, $A$ is a uniformly elliptic matrix-valued function, $\Sigma$ is a real function bounded above and below by positive constants, and $\epss_\delta$ is a complex function whose {\bf the real part takes the value $1$ and $-1$}, and the imaginary part is positive and converges to $0$ as $\delta$ goes to 0. This is motivated from a result in \cite{NicoroviciMcPhedranMilton94} and the concept of complementary  suggested in 
\cite{LaiChenZhangChanComplementary, PendryNegative, PendryRamakrishna}. 
After introducing the reflecting complementary media,  complementary media generated by reflections, we characterize $f$ for which $\|u_\delta\|_{H^1(\Omega)}$ remains bounded as $\delta$ goes to 0. For such an $f$, we also show that $u_\delta$ converges weakly  in $H^1(\Omega)$ and provide a formula to compute the limit. \end{abstract}


\section{Introduction}
Negative index materials (NIMs) were first investigated theoretically by Veselago in \cite{Veselago} and were innovated by Pendry in~\cite{PendryNegative}. The  existence of such materials was confirmed by Shelby, Smith, and Schultz in \cite{ShelbySmithSchultz} (see also \cite{SmithPadillaVierNemat-NasserSchultz}). Cloaking space and illusion optics using NIMs were discussed in \cite{LaiChenZhangChanComplementary, LaiNgChenHanXiaoZhangChanIllusion} (see also \cite{NicoroviciMcPhedranMilton94}) based on the concept of complementary and in 
 \cite{AmmariCiraoloKangLeeMilton, BouchitteSchweizer10, BrunoLintner07, KohnLu, MiltonNicorovici, Milton-folded, MiltonNicoroviciMcPhedranPodolskiy, NicoroviciMcPhedranMilton94} based on the anomalous localized resonance.  Perfect lens using NIMs was studied in \cite{MiltonNicoroviciMcPhedranPodolskiy,NicoroviciMcPhedranMilton94, PendryCylindricalLenses, PendryRamakrishna}.

\medskip 
The first motivation of this work comes from the following two dimensional result of Nicorovici et. al. in \cite{NicoroviciMcPhedranMilton94}. Let $0 < r_{1} < r_{2} < R$, and $f \in L^2(B_R)$. Here and in what follows, for $r>0$,  $B_{r}$ denotes the ball centered at 0 of radius $r$. Set $r_3 = r_2^2/ r_1$. Assume that  $R > r_3$ and $\supp f \cap \{ x \; |x| < r_3\} = \O$. Let $u_\delta \in H^1_0(B_R)$ be the unique solution to the equation
\begin{equation}\label{Milton}
\dive(\eps_\delta \nabla u_\delta) =  f \mbox{ in } B_R .    
\end{equation} 
Here 
\begin{equation*}
\eps_\delta = \eps + i 1_{r_1 < |x| < r_2}, 
\end{equation*}
where 
\begin{equation*}
\eps(x) =  \left \{\begin{array}{cl}
-1 & \mbox{ if } r_{1} < |x| < r_{2}, \\[6pt]
1  & \mbox{ otherwise}. 
\end{array} \right.
\end{equation*}
Physically, the imaginary part of $\eps_{\delta}$ is the loss of the medium. It is showed in \cite{NicoroviciMcPhedranMilton94} by separation of variables that 
\begin{equation}\label{key-point}
u_\delta \to {\cal U} \mbox{ for }  |x| > r_3,
\end{equation}
where ${\cal U} \in H^1_0(B_R)$ is the unique  solution to the equation
\begin{equation*}
\Delta {\cal U}= f \mbox{ in } B_R.
\end{equation*}
The surprising fact on this result is that \eqref{key-point} holds for {\bf any} $f$ with $\supp f \cap B_{r_3} = \O$. From \eqref{key-point}, one might say that the region $\{r_{2} < |x| < r_{3}\}$ is canceled by the one in $\{r_{1} < |x| < r_{2}\}$ and the total system is effectively equal to the free space; invisibility is achieved.

\medskip
The following questions naturally arise:  
\begin{itemize}
\item What happens if $\supp f \cap B_{r_3} \not = \O$? 
\item Is the radial symmetry necessary? If not, what are conditions on $\eps$? 
\item  What happens in the finite frequency regime? 
\item Do similar phenomena hold in three dimensions? If yes, under which conditions?   
\end{itemize}

Another motivation of this work is the concept of complementary media which was suggested in \cite{LaiChenZhangChanComplementary,PendryRamakrishna0, PendryRamakrishna} (see also \cite{NicoroviciMcPhedranMilton94, PendryNegative}). This concept has played an important role in the study of NIMs and its applications such as cloaking, perfect lens, and illusion optics  see  \cite{LaiChenZhangChanComplementary, LaiNgChenHanXiaoZhangChanIllusion, Milton-folded, PendryNegative, PendryRamakrishna}. 
Although many examples have been suggested, this concept has not be defined in a precise manner. A common point in examples studied is
\begin{equation}\label{point-common}
F_{*} a = - b \mbox{ in } D_{2}, 
\end{equation}
for some differomorphism $F: D_{1} \to D_{2}$ if  a matrix $a$ defined in a region $D_{1}$ is complementary to a matrix $b$ defined in a region $D_{2}$. 
Here\begin{equation*}
F_*a(y) = \frac{D F  (x)  a(x) D F ^{T}(x)}{J(x)} \mbox{ where }  x =F^{-1}(y) \mbox{ and } J(x) = |\det D F(x)|.
\end{equation*}
It is easy to verify in two dimensions that if  $F : \{r_{1} < |x| < r_{2}\} \to \{r_{2} < |x| < r_{3}\} $ with $r_{3} = r_{2}^{2}/ r_{1}$ defined by $F(x) = r_{2}^{2} x /|x|^{2}$ then $F$ is a diffeomorphism and 
\begin{equation*}
F*(-I) = -  I \mbox{ in }  \{r_{2} < |x| < r_{3}\}. 
\end{equation*}
In other words, the medium $-I$ in $\{r_{1} < |x| < r_{2}\} $ is complementary to the medium $I$ in $\{r_{2} < |x| < r_{3}\}$: complementary media appears in the setting of Nocorovici et. al.'s.

\medskip
In this paper, we address the above questions. For this end, we first introduce the notion of reflecting complementary media.  Similar phenomena as in \eqref{key-point} take place for media inherits this property in the  quasistatic and the finite frequency regimes. 
Two media (two matrices in two regions in the quasistatic regime) are called reflecting complementary if they are complementary and the complementary is generated by a reflection which  satisfies some mild conditions. The motivation for the definition of this notion comes from the reflection technique used in  an heuristic argument for  \eqref{key-point} in Section~\ref{sect-heuristic}.  We then establish results in the spirit of \eqref{key-point} for media of this property. The analysis again has root from the heuristic argument. The key of the analysis is the derivation of two Cauchy problems for elliptic equations  by the relection technique using the reflecting complementary property. 
Concerning the analysis, we characterize $f$ (based on the compatibility condition in Definition~\ref{compatible-k}) for which $\|u_\delta\|_{H^1}$ remains bounded as $\delta \to 0_+$; moreover,  we show that for such a function $f$, the limit of $u_\delta$ exists as $\delta \to 0_{+}$ (Theorem~\ref{thm2}). We also provide a formula to compute the limit which involves only the solutions of standard elliptic equations (no sign changing coefficients), and show that the limit has properties in the spirit of \eqref{key-point} (Theorem~\ref{thm2-1}).  To our knowledge, the results presented in this paper are new even in the 2d-quasistatic regime.   

\medskip
The use of  reflections to study NIMs has been considered previously in \cite{Milton-folded}. However, there is a big difference between the use of reflections in \cite{Milton-folded} and in this paper.  In \cite{Milton-folded}, the authors used reflections as a change of variables to obtain a  new simple setting from an old more complicated one and hence the analysis of the old problem becomes simpler.  In this paper, we use reflections to 
derive the two Cauchy problems. This derivation makes use essentially the complementary property of media. The global or non-glocal existence of the solutions to these Cauchy problems will determine the boundedness or unboundedness  of the $H^{1}$-norm of the solutions.

\medskip
The goals and the setting of  this paper are different from the ones of Ammari et al.'s  in \cite{AmmariCiraoloKangLeeMilton} and Kohn et. al.'s in \cite{KohnLu}.  In this paper, we introduce the concept of  reflecting complementary media and we study  the boundedness of $\| u_{\delta}\|_{H^{1}}$ and the limit of $u_{\delta}$ in the whole domain as $\delta \to 0$ in the quasistatic and the finite frequency regimes.  In \cite{AmmariCiraoloKangLeeMilton, KohnLu}, the authors investigate the unboundedness of $\delta^{1/2} \| u_{\delta}\|_{H^{1}}$ for piecewise constant media (up to  a diffeomorphism in \cite{KohnLu}) in the quasistatic case. 
It is clear that the boundedness of $\| u_\delta \|_{H^1}$ implies the boundedness of $\delta^{1/2}\| u_\delta \|_{H^1}$ and  the unboundedness of $\delta^{1/2}\| u_\delta \|_{H^1}$ implies the unboundedness of $\| u_\delta \|_{H^1}$.  The media considered in \cite{KohnLu} are of  the reflecting complimentary property; however  the ones in \cite{AmmariCiraoloKangLeeMilton} are, in general,  not  (the radial setting considered in \cite[Section 5]{AmmariCiraoloKangLeeMilton} is an exclusion).  In \cite{AmmariCiraoloKangLeeMilton}, the authors also deal with the boundedness of $u_{\delta}$ in some region. To make use of their results mentioned above, one needs detailed information on the spectral properties of certain boundary integral operators. This information is difficult to come by in general. The method in this paper is different from and more elementary than the spectral one in  \cite{AmmariCiraoloKangLeeMilton} and the variational one in \cite{KohnLu}. 

\medskip
The method in this paper is used in \cite{MinhLoc} and developed in \cite{Ng-Negative-Cloaking}. In \cite{MinhLoc} the authors  study the complete resonance and  localized resonance in plasmonic structures whileas in \cite{Ng-Negative-Cloaking} the author investigate the cloaking via complementary media. 

\medskip
Our paper is organized as follows. Section~\ref{sect-complementary} contains two subsections devotes to the concept of reflecting complementary media. In the first subsection, we present the heuristic argument to obtain~\eqref{key-point} and to motivate the definition of this concept. The second subsection devotes to the definition.  In Section~\ref{sect-property}, we state and prove properties on the reflecting complementary media. More precisely, we state and prove Theorems~\ref{thm2} and \ref{thm2-1}  and present their two corollaries there. 



\section{Reflecting complementary media} \label{sect-complementary}
In this section, we introduce the notion of reflecting complementary media. To motivate the definition, we first present an heuristic argument, in Section~\ref{sect-heuristic},  to obtain \eqref{key-point} based on the reflecting technique. The definition of reflecting complementary media is introduced  in Section~\ref{sect-complementary}.

\subsection{Motivation - an heuristic argument for \eqref{key-point}}\label{sect-heuristic}

In this section, we present an heuristic (elementary) argument to obtain \eqref{key-point}. This argument motivates  not only the  notion  of the reflecting complementary media but also the analysis in Section~\ref{sect-property}. In this section, we {\it assume} that $u_\delta \to u \in H^1(B_{R})$. It follows that $u \in H^1_{0}(B_{R})$ is a solution to the equation 
\begin{equation*}
\dive (\eps \nabla u ) = f \mbox{ in } B_{R}. 
\end{equation*} 
Let $F$ defined in $\{|x| < r_{2} \}$ be the Kelvin transform w.r.t.  $\partial B_{r_{2}}$,  i.e.,  
\begin{equation*}
F(x) = \frac{r_{2}^2}{|x|^{2}} x \mbox{ for } |x| < r_{2}. 
\end{equation*}
Let $u_{1}$ defined in $\{|x| > r_{2}\}$ be the Kelvin transform $F$ of $u$, i.e.,   
\begin{equation}\label{}
u_{1}(x) = u \circ F^{-1} (x) \quad \mbox{ for } |x| > r_{2}. 
\end{equation}
Then, by the transmission condition on $\partial B_{r_2}$, we have
\begin{equation}\label{transmission-1-cylind}
u_{1} = u \quad \mbox{ and } \quad \partial_{r} u_{1} \Big|_{r \to {r_{2}}_{+}}= \partial_{r} u \Big|_{r \to {r_{2}}_{+}}\quad \mbox{ for } |x| =  r_{2}. 
\end{equation}
Since $F$ is  a Kelvin transform and $\supp f \cap \{|x| < r_{3} \} = \O$, it follows  that  \begin{equation}\label{eq-u1-11}
\dive (\hat \eps \nabla u_{1}) = 0 \mbox{ for } |x| > r_{2},  
\end{equation}
where 
\begin{equation}\label{typical-example}
\hat \eps (x) =  \left \{\begin{array}{cl} 1 & \mbox{ if } r_{2 } < |x| < r_3, \\[6pt]
- 1& \mbox{ if } |x| >  r_3.
\end{array} \right.
\end{equation}
(Note that $F$ transforms $\partial B_{r_{1}}$ into $\partial B_{r_3}$.)   By the unique continuation principle, we have 
\begin{equation}\label{uu1}
u_{1} = u \mbox{ in } \{ r_{2} < |x| < r_{3}\}. 
\end{equation}
Let $G$ defined in $\{|x| > r_{3} \}$ be the Kelvin transform w.r.t. $\partial B_{r_3}$,  i.e., 
\begin{equation*}
G(x) = \frac{r_{3}^2}{|x|^{2}} x \mbox{ for } |x| > r_{3}. 
\end{equation*}
Define $u_{2}$ in $\{|x| < r_{3}\}$,  the Kelvin transform $G$ of $u_{1}$,  as follows 
\begin{equation}\label{-cylind}
u_{2} (x)= u_{1} \circ G^{-1} (x) \quad \mbox{ for } |x| < r_{3}. 
\end{equation}
Similar to \eqref{transmission-1-cylind}, we have
\begin{equation}\label{transmission-2-cylind}
u_{2} = u_{1} \quad \mbox{ and } \quad  \partial_{r} u_{2} \Big|_{r \to {r_{3}}_{-}}=  \partial_{r} u_{1} \Big|_{r \to {r_{3}}_{-}}\quad \mbox{ for } |x|  = r_{3}. 
\end{equation}
It follows from \eqref{uu1} that
\begin{equation}\label{transmission-3-cylind}
u_{2} = u \quad \mbox{ and } \quad  \partial_{r}   u_{2} \Big|_{r \to {r_{3}}_{-}} =  \partial_{r} u  \quad \mbox{ for } |x|  = r_{3}. 
\end{equation}
We also have 
\begin{equation}\label{transmission-4-cylind}
\Delta u_{2} = 0 \mbox{ in } |x| < r_{3}, 
\end{equation}
by the property of the Kelvin transforms.  Define ${\cal U}$ by
\begin{equation}\label{def-U-P}
{\cal U} (x)=  \left \{\begin{array}{cl} u (x) & \mbox{ if } |x| > r_{3}, \\[6pt]
u_{2} (x), & \mbox{ if } |x| <  r_{3}.  
\end{array} \right.
\end{equation}
Since $\Delta u = f$ for $|x| > r_{3}$, it follows from \eqref{transmission-3-cylind} and \eqref{transmission-4-cylind} that 
\begin{equation*}
\Delta {\cal U} = f \mbox{ in } B_{R}. 
\end{equation*}
Therefore, we obtain \eqref{key-point}.  

\subsection{Reflecting complementary media}

In this section, we introduce the notion of reflecting complementary media.  Let 
$\Omega_{1} \subset \subset \Omega_{2} \subset \subset \Omega_{3} \subset \subset \Omega$ be smooth connected bounded open subsets of $\mR^{d}$ ($d = 2, \, 3$). 
Let $A$ be a measurable matrix function and $\Sigma$ be a measurable real function defined in $\Omega$ such that
\begin{equation}\label{pro-A}
\frac{1}{\Lambda} |\xi|^2 \le \langle A(x) \xi, \xi \rangle \le \Lambda |\xi|^2 \quad \forall \, \xi \in \mR^d,
\end{equation}
for a.e. $x \in \Omega$ and  for some $0< \Lambda < + \infty$ and
\begin{equation}\label{pro-Sigma}
0 < \mathop{\ess \inf}_{\Omega} \Sigma \le \mathop{\ess sup}_{\Omega} \Sigma < +\infty. 
\end{equation}

\medskip
Set 
\begin{equation}\label{def-eDelta}
\epss_\delta (x) = \left\{\begin{array}{cl} -1 + i \delta & \mbox{ if }  x \in \Omega_2 \setminus \Omega_1,  \\[6pt]
1 & \mbox{ otherwise}.
\end{array}\right.
\end{equation}

We are later interested in  the behavior of the unique solution $u_\delta \in H^1_0(\Omega)$  to the equation
\begin{equation}\label{eq-uDelta}
\dive (\epss_\delta A \nabla u_\delta) + k^{2} \epss_{0} \Sigma u_{\delta} = \epss_0 f  \mbox{ in } \Omega,
\end{equation}
as $\delta \to 0$. 

\medskip
We are ready to give

\begin{definition}[Reflecting complementary media] \label{def-Geo} The media $(A, \Sigma)$ in $\Omega_{3} \setminus \Omega_{2}$ and $(-A, - \Sigma)$ in $\Omega_{2} \setminus \Omega_{1}$  are said to be  reflecting complementary if 
there exists a diffeomorphism $F: \Omega_{2} \setminus \bar \Omega_{1} \to \Omega_{3} \setminus \bar \Omega_{2}$ such that 
\begin{equation}\label{cond-ASigma}
F_*A (x) = A(x), \quad  F_{*} \Sigma (x) =   \Sigma(x)  \mbox{ for  } x \in  \Omega_3 \setminus \bar\Omega_2, 
\end{equation}
\begin{equation}\label{cond-F-boundary}
F(x) = x \mbox{ on } \partial \Omega_2. 
\end{equation}
and the following two conditions hold:
\begin{enumerate}
\item[1.] There exists an diffeomorphism extension of $F$, which is still denoted by  $F$, from $\Omega_{2} \setminus \{x_{1}\} \to \Omega_{4} \setminus \bar \Omega_{2}$ for some $x_{1} \in \Omega_{1}$ and some smooth open subset $\Omega_{4}$ of $\mR^{d}$ with  $\Omega_{3} \subset \Omega_{4}$. 

\item[2.] There exists a diffeomorphism $G: \Omega_{4} \setminus \bar \Omega_{3} \to \Omega_{3} \setminus x_{2}$ for some $x_{2} \in \Omega_{3}$  such that  \footnote{In \eqref{cond-F-boundary} and \eqref{cond-G-boundary}, $F$ and $G$ denote some  diffeomorphism extensions of $F$ and $G$ in a neighborhood of $\partial \Omega_2$ and of $\partial \Omega_3$.}
\begin{equation}\label{cond-G-boundary}\
\quad G(x) = x \mbox{ on } \partial \Omega_3,
\end{equation}
and
\begin{equation}\label{extension}
G \circ F : \Omega_1  \to \Omega_3 \mbox{ is a diffeomorphism if one sets } G\circ F(x_1) = x_2.
\end{equation}
\end{enumerate} 
\end{definition}


\medskip








Here and in what follows, we use the standard notations:
\begin{equation}\label{def-*}
{\cal F}_*{\cal A}(y) = \frac{D {\cal F} (x) {\cal A}(x) D {\cal F}^{T}(x)}{J(x)}, \;\;\; {\cal F}_* {\it \Sigma}(y) = \frac{{\it \Sigma}(x)}{J(x)}, \;\;\; \mbox{and} \;\;\; {\cal F}_* \textsl{f}(y) = \frac{ \textsl{f} (x)}{J(x)},
\end{equation}
where $x = {\cal F}^{-1}(y)$ and $J(x) = |\det D {\cal F}(x)|$.


\medskip

Some comments on the definition are: 
\begin{enumerate}
\item[i)] If $k=0$, then  the condition on $\Sigma$ is irrelevant in Definition~\ref{def-Geo}.


\item[ii)] Condition \eqref{cond-ASigma} implies that $(A, \Sigma)$ in $\Omega_{3} \setminus \Omega_{2}$ and $(-A, - \Sigma)$ in $\Omega_{2} \setminus \Omega_{1}$ are complementary in the usual sense. The term ``reflecting'' in the definition comes from \eqref{cond-F-boundary} and the assumption $\Omega_{1} \subset \Omega_{2} \subset \Omega_{3}$. Conditions  \eqref{cond-ASigma} and \eqref{cond-F-boundary} are the main assumptions in the definition. 

\item[iii)]
Condition~\eqref{cond-ASigma} makes sure that $u$ (the ``solution'' for $\delta =0$) and $u_{1} : = u \circ F$ satisfy the same equation in $\Omega_{3} \setminus \Omega_{2}$; hence the reflecting technique in Section~\ref{sect-heuristic} can be used. 
Condition \eqref{cond-F-boundary} and \eqref{cond-G-boundary}  assure that $u = u_1$ on $\partial \Omega_2$ and $u_2 = u_1$ on $\partial \Omega_3$ where $u_2 = u_1 \circ G^{-1}$. Condition \eqref{extension} is a technical one which is required by the proof.

\item[iv)] Conditions 1) and 2) in the definition are  mild assumptions. Introducing $G$ in the definition makes the analysis more accessible as in Section~\ref{sect-heuristic} (see  Remark~\ref{rem-compatibility} for other comments on $G$).

\item[v)] In general, it is difficult to verify whether \eqref{cond-ASigma} holds for some $F$. In practice, 
to obtain the reflecting complementary in $\Omega_{2} \setminus \Omega_{1}$ for $(A, \Sigma)$ in $\Omega_{3} \setminus \Omega_{2}$, it suffices to choose a diffeomorphism $F: \Omega_2 \setminus \{x_1\} \to \Omega_4 \setminus \bar \Omega_2$ for some $x_{1} \in \Omega_{1}$ and for some smooth bounded open subset $\Omega_{4}$ containing $\Omega_{3}$, and define $(-A, -\Sigma)$ in $\Omega_{2} \setminus \Omega_{1}$ by  $(-F^{-1}_{*} A, -F^{-1}_{*} \Sigma)$. 

\end{enumerate}

It is clear that the medium $\eps$ given in \eqref{typical-example} has the reflecting complementary property with $\Omega_i = B_{r_i}$ (for $i=1, 2, 3$) and $\Omega = B_R$ where $r_4 : = +\infty$, and $F$  and $G$ are the Kelvin transforms w.r.t. $\partial B_{r_{2}}$ and  $\partial B_{r_{3}}$ resp. 

\begin{remark}
Concerning reflecting complementary media, the 2d quasistatic case is quite special in the sense that two constant media (two constant matrices) can be complementary (see the example in the introduction). In fact, in the 2d finite frequency case, it seems that there do not exist two constant media (two constant matrices and two constant functions) which are complementary and in  the 3d case  there do not exist two constant media (two constant matrices) which are complementary.  
\end{remark}




\section{Properties related to reflecting complementary media}\label{sect-property}

\subsection{Statement of the results} \label{sect-Sta-fre}

 Let $\Omega_{1} \subset \subset \Omega_{2} \subset \subset \Omega_{3} \subset \subset \Omega$ be smooth connected bounded open subsets of $\mR^{d}$ ($d = 2, \, 3$). 
 Let $A$ be a measurable matrix function and $\Sigma$ be a measurable real function defined in $\Omega$ such that
\begin{equation}\label{pro-A}
\frac{1}{\Lambda} |\xi|^2 \le \langle A(x) \xi, \xi \rangle \le \Lambda |\xi|^2 \quad \forall \, \xi \in \mR^d,
\end{equation}
for a.e. $x \in \Omega$ and  for some $0< \Lambda < + \infty$, and
\begin{equation}\label{pro-Sigma}
0 < \mathop{\ess \inf}_{\Omega} \Sigma \le \mathop{\ess sup}_{\Omega} \Sigma < +\infty. 
\end{equation}
We will assume that $A$ is piecewise differentiable  in $\Omega$ in three dimensions \footnote{This condition is necessary to obtain the uniqueness for the Cauchy problems.}. 

\medskip
Set 
\begin{equation}\label{def-eDelta}
\epss_\delta (x) = \left\{\begin{array}{cl} -1 + i \delta & \mbox{ if }  x \in \Omega_2 \setminus \Omega_1,  \\[6pt]
1 & \mbox{ otherwise}.
\end{array}\right.
\end{equation}
We are  interested in  the behavior of the unique solution $u_\delta \in H^1_0(\Omega)$  to the equation
\begin{equation}\label{eq-uDelta}
\dive (\epss_\delta A \nabla u_\delta) + k^{2} \epss_{0} \Sigma u_{\delta} = \epss_0 f  \mbox{ in } \Omega,
\end{equation}
as $\delta \to 0$. 

\medskip
Throughout this section, we will assume that:
\begin{equation}\label{toto11}
\mbox{Systems \eqref{frequency-assumption} and \eqref{frequency-assumption-2} have only zero solutions in $H^1(\Omega \setminus \bar \Omega_2)$ and $H^1(\Omega)$ resp.}, 
\end{equation}
where
\begin{equation}\label{frequency-assumption}\left\{
\begin{array}{cl}
\dive (A \nabla v) + k^2 \Sigma v = 0 & \mbox{ in } \Omega \setminus \bar \Omega_2,\\[6pt]
v = 0 & \mbox{ on } \partial \Omega \cup \partial \Omega_{2}, 
\end{array}\right.
\end{equation}
and
\begin{equation}\label{frequency-assumption-2}\left\{
\begin{array}{cl}
\dive (\hat A \nabla U) + k^2 \hat \Sigma U = 0 & \mbox{ in } \Omega, \\[6pt]
U = 0 & \mbox{ on } \partial \Omega. 
\end{array} \right.
\end{equation}

Here $\hat A$ and $\hat \Sigma$ are defined as follows 
\begin{equation}\label{def-hatA-hatSigma}
\hat A := \left\{ \begin{array}{cl}
A & \mbox{ if } x \in \Omega \setminus \Omega_3,\\[6pt]
G_*F_*A & \mbox{ if } x \in \Omega_3,
\end{array}
\right.
\quad \mbox{ and } \quad  \hat \Sigma := \left\{ \begin{array}{cl}
\Sigma & \mbox{ if } x \in \Omega \setminus \Omega_3,\\[6pt]
G_*F_*\Sigma & \mbox{ if } x \in \Omega_3. 
\end{array}
\right.
\end{equation}
We also define
\begin{equation}\label{def-hatf}
\hat f : = \left\{ \begin{array}{cl}
f & \mbox{ if } x \in \Omega \setminus \Omega_3,\\[6pt]
G_*F_*f & \mbox{ if } x \in \Omega_3.
\end{array}
\right.
\end{equation}

\begin{remark}
The well-posedness of  \eqref{frequency-assumption} and \eqref{frequency-assumption-2} always hold for $k=0$. In the case $k>0$,  if one is interested the corresponding problem on the whole space in which  outgoing solutions are considered, the well-posedness assumption is not necessary. 
\end{remark}


In what follows we assume that $k > 0$. The case $k=0$ is similar and even easier to obtain.  The first main result in this section is

\begin{theorem}\label{thm2} Let $d=2, \, 3$, $\delta>0$, $f \in L^2(\Omega)$ and let $u_\delta \in H^1_0(\Omega)$ be the unique solution to equation \eqref{eq-uDelta}:
\begin{equation*}
\dive (\epss_\delta A \nabla u_\delta) + k^{2} \epss_{0} \Sigma u_{\delta} = \epss_0 f  \mbox{ in } \Omega.
\end{equation*}
Assume that the media $(A, \Sigma)$ in $\Omega_{3} \setminus \Omega_{2}$ and  $(-A, - \Sigma)$ in $\Omega_{2} \setminus \Omega_{1}$  are  reflecting complementary. We have
\begin{enumerate}
\item[a)] Case 1: $f$ is compatible with the medium. Then  $(u_\delta)$ converges weakly in $H^1(\Omega)$ and strongly in $L^2(\Omega)$ to $u_0 \in H^1_0(\Omega)$, the unique solution to the equation
\begin{equation}\label{limit-eq-k}
\dive (\epss_0 \nabla u_0) + k^2 \epss_0 \Sigma u_0 = \epss_0 f \mbox{ in } \Omega.
\end{equation}
as $\delta \to 0$. 
\item[b)] Case 2: $f$ is not compatible with the medium. We have
\begin{equation}\label{limit-energy-k}
\lim_{\delta \to 0}\| u_\delta \|_{H^1(\Omega)} = + \infty.
\end{equation}
\end{enumerate}
\end{theorem}

In the statement of Theorem~\ref{thm2}, we use the following

\begin{definition}[Compatibility condition]\label{compatible-k} Assume that the media $(A, \Sigma)$ in $\Omega_{3} \setminus \Omega_{2}$ and  $(-A, - \Sigma)$ in $\Omega_{2} \setminus \Omega_{1}$  are  reflecting complementary.  Then $f \in L^2(\Omega)$ is said to be compatible with the system if and only if  there exist
$U \in H^1(\Omega_3 \setminus \Omega_2)$ and $V \in H^1(\Omega_3 \setminus \Omega_2)$  such that
\begin{equation}\label{def-U-fre}\left\{
\begin{array}{cl}
\dive (A \nabla U) + k^2 \Sigma U = F_*f - f & \mbox{ in } \Omega_3 \setminus \bar \Omega_2, \\[6pt]
U = 0 & \mbox{ on } \partial \Omega_2, \\[6pt]
A \nabla U \cdot \eta   = 0 & \mbox{ on } \partial \Omega_2,
\end{array} \right.
\end{equation}
and
\begin{equation}\label{def-V-fre}
\left\{
\begin{array}{cl}
\dive (A \nabla V)  + k^2 \Sigma V = f  & \mbox{ in } \Omega_3 \setminus \bar \Omega_2, \\[6pt]
V = W \Big|_{\mathrm{ext}}& \mbox{ on } \partial \Omega_3, \\[6pt]
A \nabla V \cdot \eta   = A \nabla W \cdot \eta \Big|_{\mathrm{ext}}  & \mbox{ on } \partial \Omega_3.
\end{array} \right.
\end{equation}
Here  $W \in H^1(\Omega \setminus \partial \Omega_3)$ is the unique solution to the system
\begin{equation}\label{def-W-fre}\left\{
\begin{array}{cl}
\dive (\hat A \nabla  W) + k^2 \hat \Sigma W = \hat f & \mbox{ in } \Omega \setminus \partial \Omega_3, \\[6pt]
W = 0 & \mbox{ on } \partial \Omega, \\[6pt]
[W] = - U & \mbox{ on } \partial \Omega_3, \\[6pt]
[\hat A \nabla W \cdot \eta ] = - A \nabla U \cdot \eta & \mbox{ on } \partial \Omega_3.
\end{array} \right.
\end{equation}
\end{definition}

The compatibility condition is an intrinsic one, i.e., it does not depend on the choice of $F$ and $G$. In fact, 
it is equivalent to the boundedness of $\| u_{\delta}\|_{H^{1}(\Omega)}$ as $\delta \to 0$. Given $F$, there are infinitely many choices of $G$. A choice  of $G$ that would make the compatibility condition more accessible is preferred.  Problems \eqref{def-U-fre} and \eqref{def-V-fre} are the two Cauchy problems obtained from the reflecting technique. 
The reflecting complementary condition is necessary so that $(\|u_{\delta \|_{H^{1}(\Omega)}})$ explodes for some $f$ (see \cite{KohnLu}).  

\medskip
The uniqueness of $U$ and $V$ follow from the unique continuation principle (see,  e.g., \cite{AlessandriniUnique, Protter60}). The existence and uniqueness of $W$ are standard and follow from Fredholm's theory (see,  e.g.,  \cite{BrAnalyse}) since  system \eqref{frequency-assumption-2} is well-posed.  



\medskip
The next result is in the spirit of \eqref{key-point}. 

\begin{theorem}\label{thm2-1} Let $d=2, \, 3$, $\delta>0$, $f \in L^2(\Omega)$ and let $u_\delta \in H^1_0(\Omega)$ be the unique solution to equation \eqref{eq-uDelta}:
\begin{equation*}
\dive (\epss_\delta A \nabla u_\delta) + k^{2} \epss_{0} \Sigma u_{\delta} = \epss_0 f  \mbox{ in } \Omega.
\end{equation*}
Assume that the media $(A, \Sigma)$ in $\Omega_{3} \setminus \Omega_{2}$ and  $(-A, - \Sigma)$ in $\Omega_{2} \setminus \Omega_{1}$  are  reflecting complementary and $f$ is compatible with the system. Then $(u_\delta)$ converges weakly to $NI(f)$ in $H^{1}(\Omega)$ where 
\begin{equation}\label{def-NI}
NI(f)=
\left\{\begin{array}{cl}
W & \mbox{ if } x \in \Omega \setminus \Omega_3, \\[6pt]
V  & \mbox{ if } x \in \Omega_3 \setminus \Omega_2, \\[6pt]
(U + V) \circ F & \mbox{ if } x \in \Omega_2 \setminus \Omega_1, \\[6pt]
W \circ G \circ F & \mbox{ if } x \in \Omega_1.
\end{array}\right.
\end{equation}
\end{theorem}

Here $U, V$, and $W$ are given in Definition~\ref{compatible-k}. 

\medskip  If 
$f = 0$ in $\Omega_{3}$ then $U = 0$. In this case, the compatibility condition is  equivalent to the existence of $V \in H^{1}(\Omega_{3} \setminus  \bar \Omega_{2})$ 
to the Cauchy problem
\begin{equation}\label{VVV}
\left\{\begin{array}{cl}
\dive (A \nabla V) + k^{2} \Sigma V = 0 & \mbox{ in } \Omega_3 \setminus \bar \Omega_2, \\[6pt]
V = W \Big|_{\mathrm{ext}} & \mbox{ on } \partial \Omega_3, \\[6pt]
A \nabla V \cdot \eta   = A \nabla W \cdot \eta \Big|_{\mathrm{ext}}  & \mbox{ on } \partial \Omega_3, 
\end{array} \right.
\end{equation}
where $W \in H^{1}_{0}(\Omega)$ is the unique solution to
\begin{equation*}
\dive (\hat A \nabla W) + k^{2} \hat \Sigma W= f \mbox{ in } \Omega. 
\end{equation*}

We have 

\begin{corollary}\label{cor1} Let $d=2, \, 3$, $f \in L^2(\Omega)$ with $\supp f \cap \Omega_{3} = \O$.  Assume that the media $(A, \Sigma)$ in $\Omega_{3} \setminus \Omega_{2}$ and  $(-A, - \Sigma)$ in $\Omega_{2} \setminus \Omega_{1}$  are  reflecting complementary and there exists a solution  $V \in H^{1}(\Omega_{3} \setminus \bar \Omega_{2})$ to \eqref{VVV}.  Then $f$ is compatible to the system. 

\end{corollary}

It follows from Corollary~\ref{cor1} that if $\supp f \cap \Omega_{3} = \O$ and $G_{*} F_{*} A(x) = A(x)$ and  $G_{*} F_{*} \Sigma(x) = \Sigma(x)$ for $x \in \Omega_{3} \setminus \bar \Omega_{2}$ then $V = W$. We obtain

\begin{corollary} \label{cor2}  Let $d=2, \, 3$, $f \in L^2(\Omega)$ with $\supp f \cap \Omega_{3} = \O$.  Assume that the media $(A, \Sigma)$ in $\Omega_{3} \setminus \Omega_{2}$ and  $(-A, - \Sigma)$ in $\Omega_{2} \setminus \Omega_{1}$  are  reflecting complementary and \begin{equation}\label{compa-1}
G_{*} F_{*} A(x) = A(x) \quad \mbox{ and } \quad G_{*} F_{*} \Sigma(x) = \Sigma(x) \quad  \mbox{ for } \quad x \in \Omega_{3} \setminus \bar\Omega_{2}. 
\end{equation}
Then $f$ is compatible to the system. 
\end{corollary} 

\begin{remark} It is easy to verify that the $2d$ setting considered in the introduction satisfies the assumptions of Corollary~\ref{cor2}. 
\end{remark}

\begin{remark} In this paper, we characterize the behavior of $u_{\delta}$ as $\delta \to 0$ for compatible $f$. The paper \cite{Ng-Negative-Cloaking}  develops the method introduced here to deal with this problem without the assumption on the compatibility in the cloaking setting. 
\end{remark}

\subsection{ Proofs of Theorems~\ref{thm2} and \ref{thm2-1}} \label{proof-thm2}

This section containing two subsections is devoted to the proof of Theorems~\ref{thm2} and \ref{thm2-1}. In the first subsection, we establish basis properties of solutions to  equation \eqref{eq-uDelta} such as the existence, uniqueness, and stability, and establish a result on the  change of variables concerning reflections.  The proof of Theorem \ref{thm2} and \ref{thm2-1} are given in the second subsection.

\subsubsection{Preliminaries}\label{sec-frequency-pre}

This section contains two lemmas. The first one is  on the wellposedness of \eqref{eq-uDelta}.  

\begin{lemma}\label{lem1} Let $d=2, \, 3$, $k>0$,  $0 < \delta < 1$, $g \in H^{-1}(\Omega)$ $($the duality of $H^1_0(\Omega)$$)$ and let $\epss_{\delta}$ be defined in \eqref{def-eDelta}. 
Assume that $A$ and $\Sigma$ satisfy \eqref{pro-A} and \eqref{pro-Sigma}, and \eqref{toto11} holds. Then there exists a unique solution $v_\delta \in H^1_0(\Omega)$ to the equation
\begin{equation}\label{eqd}
\dive (\epss_\delta A \nabla v_\delta) + k^2 \epss_0 \Sigma v_\delta = g  \mbox{ in } \Omega.
\end{equation}
Moreover,
\begin{equation}\label{stability}
\| v_\delta \|_{H^1(\Omega)} \le  C \Big(\frac{1}{\delta} \| g \|_{H^{-1}(\Omega)}+ \| g\|_{L^2(\Omega_1)} + \| g\|_{L^2(\Omega_2 \setminus \bar \Omega_1)}  + \| g\|_{L^2(\Omega \setminus \bar \Omega_2)} \Big),
\end{equation}
for some positive constant $C$ independent of $g$ and $\delta$, as $\delta$ is small.
\end{lemma}

\noindent{\bf Proof.} The existence of $v_\delta$ follows from the uniqueness of $v_{\delta}$ by Fredholm's theorem (see, e.g.,  \cite{BrAnalyse}). We now establish the uniqueness of $v_{\delta}$ by showing that $v_\delta = 0$ if $v_\delta \in H^1_0(\Omega)$ is a solution to the equation
\begin{equation*}
\dive(\epss_\delta A \nabla v_\delta) + k^2 \epss_0 \Sigma v_\delta = 0 \mbox{ in } \Omega.
\end{equation*}
Multiplying the above equation by $\bar v_\delta$ (the conjugate of $v_{\delta}$) and integrating the obtained expression on $\Omega$, we have
\begin{equation*}
\int_{\Omega} \epss_\delta \langle A \nabla v_\delta, \nabla v_\delta \rangle \, dx - \int_{\Omega} k^2 \epss_0 \Sigma |v_\delta|^2 \, dx = 0.
\end{equation*}
This implies, by considering the imaginary part,
\begin{equation*}
\int_{\Omega_2 \setminus \Omega_1} \langle A \nabla v_\delta, \nabla v_\delta \rangle \, dx= 0.
\end{equation*}
It follows from  \eqref{pro-A} that $v_\delta$ is constant in $\Omega_2 \setminus \Omega_1$. Thus $v_\delta = 0$ in $\Omega_2 \setminus \Omega_1$ since
$\dive(\epss_\delta A \nabla v_\delta) + k^2 \epss_0 \Sigma v_\delta = 0$ in $\Omega_2 \setminus \Omega_1$.
This implies $v_\delta = 0 $ in $\Omega \setminus \Omega_{2}$ and in $\Omega_{1}$ by \eqref{toto11}. Here to establish $v_{\delta} = 0 $ in $\Omega_{1}$, we considered the function $V_{\delta}$ defined in $\Omega$ as follows $V_{\delta} = v_{\delta} \circ  F^{-1} \circ G^{-1}$ in $\Omega_{3}$ and $V_{\delta} = 0$ in $\Omega \setminus \Omega_{3}$ and used the well-posedness of  \eqref{frequency-assumption-2}.  The proof of the uniqueness is complete.

\medskip
We next establish \eqref{stability} by a contradiction argument. Assume that \eqref{stability} is not true. Then 
there exists $(g_\delta) \subset H^{-1}(\Omega)$ such that
\begin{equation}\label{contradict-assumption}
\| v_\delta\|_{H^1(\Omega)} =1 \mbox{ and } \frac{1}{\delta} \|g_\delta\|_{H^{-1}} + \| g_\delta\|_{L^2(\Omega_1)} + \| g_\delta\|_{L^2(\Omega_2 \setminus \bar \Omega_1)} + \| g\|_{L^2(\Omega \setminus \bar \Omega_2)} \to 0,
\end{equation}
as $\delta \to 0$, where $v_\delta \in H^1_0(\Omega)$ is the unique solution to the equation
\begin{equation}\label{eq-v-delta}
\dive (\epss_\delta A \nabla v_\delta) + k^2 \epss_0 \Sigma v_\delta = g_\delta  \mbox{ in } \Omega.
\end{equation}
Multiplying this equation by $\bar v_\delta$ and integrating the obtained expression on $\Omega$, we have
\begin{equation*}
\int_{\Omega} \epss_\delta \langle A \nabla v_\delta, \nabla v_\delta \rangle  \, dx -  \int_{\Omega} k^2 \epss_0 \Sigma |v_\delta|^2 \, dx = - \int_\Omega g_\delta \bar v_\delta \, dx.
\end{equation*}
Considering the imaginary part and using the fact that
\begin{equation*}
\frac{1}{\delta}\Big| \int_\Omega g_\delta \bar v_\delta \Big| \le \frac{1}{\delta}\| g_\delta \|_{H^{-1}} \| v_\delta\|_{H^1(\Omega)} \to 0 \mbox{ as $\delta \to 0$ by }\eqref{contradict-assumption},
\end{equation*}
we obtain, by \eqref{pro-A},
\begin{equation}\label{toto1}
\| \nabla v_\delta\|_{L^2(\Omega_2  \setminus \Omega_1)} \to 0 \mbox{ as } \delta \to 0.
\end{equation}
Since $\dive(A \nabla v_{\delta}) + k^{2} \Sigma v_{\delta} = g_{\delta}$ in $\Omega_{2} \setminus \bar \Omega_{1}$, it follows from \eqref{contradict-assumption} and  a standard compactness argument that 
\begin{equation}\label{toto1-1}
\| v_{\delta}\|_{L^{2}(\Omega_{2} \setminus \Omega_{1})} \to 0 \mbox{ as } \delta \to 0.   
\end{equation}
A combination of \eqref{toto1} and \eqref{toto1-1} yields
\begin{equation}\label{part-1}
\| v_{\delta}\|_{H^{1}(\Omega_{2} \setminus \Omega_{1})} \to 0  \mbox{ as } \delta \to 0. 
\end{equation}
In particular, 
\begin{equation*}
\| v_{\delta}\|_{H^{1/2}(\partial \Omega_{2})}  + \| v_{\delta}\|_{H^{1/2}(\partial \Omega_{1})} \to 0  \mbox{ as } \delta \to 0.  
\end{equation*}
We derive from the well-posedness of \eqref{frequency-assumption} and \eqref{frequency-assumption-2} that 
\begin{equation}\label{part-2}
\| v_{\delta}\|_{H^{1}(\Omega \setminus \Omega_{2})} \to 0, 
\end{equation}
and 
\begin{equation}\label{part-3}
\| v_{\delta}\|_{H^{1}(\Omega \setminus \Omega_{1})} \to 0. 
\end{equation}
A combination of \eqref{part-1}, \eqref{part-2}, and \eqref{part-3} yields 
\begin{equation*}
\| v_{\delta}\|_{H^{1}(\Omega)} \to 0. 
\end{equation*}
We have a contradiction by \eqref{contradict-assumption}.  The proof of \eqref{stability} completes. \proofend


\medskip
The second lemma is on the change of variables for reflections. 

\begin{lemma}\label{lem-TO} Let $k \ge 0$, $D_1$ and $D_2$ be two smooth open subsets of $\mR^d$, $T$ be a diffeomorphism from $D_1$ onto $D_2$,  $a \in [L^\infty(D_1)]^{d \times d}$ be a matrix function, and $\sigma \in L^\infty(D_1)$ be a complex function. Fix $u \in H^1(D_1)$ and set $v = u \circ T^{-1}$. Then
\begin{equation*}
\dive (a \nabla u) + k^2 \sigma u = f \mbox{ in } D_{1}
\end{equation*}
iff
\begin{equation*}
\dive (T_*a \nabla v) + k^2 T_*\sigma v = T_* f \mbox{ in } D_{2}. 
\end{equation*}
Assume that $\Gamma_1$ and $\Gamma_2$ are open subsets of $\partial D_1$ and $\partial D_2$  such that $\Gamma_1$ and $\Gamma_2$ are smooth, $\Gamma_2 = T(\Gamma_1)$, and ${\bf T}: = T\Big|_{ \Gamma_1} \Gamma_1 \to \Gamma_2$ is a diffeomorphism \footnote{We assume here that there is an extension of $T$ in a neighborhood of $\partial D_{1}$ (which is also called $T$) such that it is a diffeomorphism.}. We have
\begin{equation*}
a \nabla u \cdot \eta_1 = g_1 \mbox{ on } \Gamma_1
\end{equation*}
iff
\begin{equation*}
T_*a \nabla v \cdot \eta_2 = g_2 \mbox{ on } \Gamma_2,
\end{equation*}
where \footnote{In the identity below, $\nabla {\bf T}$ stands for the gradient of a transformation from a $(d-1)$-manifold into a $(d-1)$-manifold, and $\det \nabla {\bf T}$ denotes the determinant of $(d-1) \times (d-1)$ matrix.}
\begin{equation*}
g_2(y) = g_1(x)/  |\det \nabla {\bf T}(x)| \mbox{ with } x = {\bf T}^{-1}(y).
\end{equation*}
Here $\eta_1$ and $\eta_2$ are the normal unit vectors on $\Gamma_1$ and $\Gamma_2$ directed to the exterior of $D_1$ and $D_2$. In particular, if
$\Gamma_1 = \Gamma_2$, ${\bf T}(x) = x$ on $\Gamma_1$, $D_2 \cap D_1 = \O$. We have
\begin{equation}\label{reflextion}
T_*a \nabla v \cdot \eta_1 = - a \nabla u \cdot \eta_1  \mbox{ on } \Gamma_1 = \Gamma_2.
\end{equation}
\end{lemma}

\noindent {\bf Proof.} Lemma \ref{lem-TO} is a consequence of the change of variables. The first equivalence relation is known and can be proved by using the weak formula for $u$ and $v$. The second equivalence follows similarly. The details are left to the reader. \proofend

\subsubsection{Proof of the first statement of  Theorem~\ref{thm2} and Theorem~\ref{thm2-1}} \label{sec-thm2}

In this section,  $f$ is compatible. The proof is derived from the following steps:

\medskip

Step 1: Let $v \in H^1_0(\Omega)$ be a solution to the equation
\begin{equation}\label{eq-lim0}
\dive(\epss_0 A \nabla v) + k^{2} \epss_{0} \Sigma v = \epss_0 f \mbox{ in } \Omega.
\end{equation}
We prove that $v = NI(f)$.

Step 2: Define $u_0: = NI(f)$. We prove that $u_0 \in H^1_0(\Omega)$ is a solution to the equation
\begin{equation*}
\dive(\epss_0 A \nabla u_0)  + k^{2} \epss_{0} \Sigma u_{0} = \epss_0 f \mbox{ in } \Omega.
\end{equation*}

Step 3: We prove that  $(u_\delta)_{0 < \delta < 1}$  is bounded in $H^1(\Omega)$.

Step 4: We prove that  $(u_\delta)$ converges weakly in $H^1(\Omega)$ and strongly in $L^2(\Omega)$ to $u_0$ as $\delta$ goes to 0.

\medskip
It is clear that the proof of the first statement of Theorem~\ref{thm2} and Theorem~\ref{thm2-1} is complete after these four steps.  We now process these steps.

\medskip
\underline{Step 1:} Assume that $v \in H^1_0(\Omega)$ is a solution to the equation
\begin{equation*}
\dive(\epss_0 A \nabla v) + k^{2} \epss_{0} \Sigma v= \epss_0 f \mbox{ in } \Omega.
\end{equation*}
Set
\begin{equation}\label{def-v1}
v_1 = v \circ F^{-1} \mbox{ in } \Omega_4 \setminus \bar \Omega_2
\end{equation}
and
\begin{equation}\label{hatepss0}
\hat \epss_0 = \left\{\begin{array}{cl} 1 & \mbox{ if } x \in \Omega_3 \setminus \Omega_2, \\[6pt]
-1 & \mbox{ if } x \in \Omega_4 \setminus \Omega_3.
\end{array}\right.
\end{equation}
Then  $v_1 \in H^1(\Omega_3 \setminus \bar \Omega_2) \cap H^1_{\loc}(\Omega_4 \setminus \bar \Omega_2)$ and,   by Lemma~\ref{lem-TO}, $v_1$ satisfies the equation
\begin{equation}\label{eq-v1}
\dive (\hat \epss_0 F_*A \nabla v_1) + k^{2}  \hat \epss_{0} F_{*}\Sigma v_{1} = \hat \epss_0 F_*f \mbox{ in } \Omega_4 \setminus \bar \Omega_2,
\end{equation}
and
\begin{equation*}
v_1 = v \mbox{ on  }  \partial \Omega_2 \quad \mbox{ and } \quad F_*A \nabla v_1 \cdot  \eta  =  A \nabla v_ \cdot \eta \Big|_{\mathrm{ext}} \mbox{ on } \partial \Omega_2.
\end{equation*}
In the last identity, we use the fact that $F_*A \nabla v_1 \cdot  \eta  = -  A \nabla v \cdot \eta \Big|_{\mathrm{int}}$ on  $\partial \Omega_2$ by \eqref{reflextion}, and $A \nabla v \cdot  \eta \Big|_{\mathrm{ext}} = - A \nabla v \cdot  \eta \Big|_{\mathrm{int}} $ on $\partial \Omega_2$ by the transmission condition on $\partial \Omega_2$.  Define
\begin{equation}\label{def-bfU}
{\bf U} = v_1 - v \mbox{ in }  \Omega_3 \setminus \bar \Omega_2.
\end{equation}
Since $F_*A = A$ and $F_{*} \Sigma = \Sigma$ in $\Omega_3 \setminus \bar \Omega_2$, it follows that
\begin{equation}\label{req-U}\left\{
\begin{array}{cl}
\dive (A \nabla {\bf U}) + k^{2} \Sigma {\bf U} = F_*f - f & \mbox{ in } \Omega_3 \setminus \bar \Omega_2, \\[6pt]
{\bf U} = 0 & \mbox{ on } \partial \Omega_2, \\[6pt]
A \nabla {\bf U} \cdot \eta   = 0 & \mbox{ on } \partial \Omega_2.
\end{array} \right.
\end{equation}
Applying the unique continuation principle (see,  e.g., \cite{AlessandriniUnique, Protter60}), from \eqref{def-U-fre}, we have
\begin{equation}\label{bfU-U}
{\bf U} = U \mbox{ in } \Omega_3 \setminus \bar \Omega_2.
\end{equation}
Define $v_2$ in $\Omega$ as follows
\begin{equation}\label{def-v2}
v_2(x) =
\left\{\begin{array}{cl}
v_1 \circ G^{-1} (x) & \mbox{ if } x \in \Omega_3, \\[6pt]
v (x) & \mbox{ if } x \in \Omega \setminus \Omega_3.
\end{array}\right.
\end{equation}
Using  \eqref{extension} and applying Lemma~\ref{lem-TO}, we have
\begin{equation}\label{eq-v2}
\dive (\hat A \nabla v_2) + k^{2 } \hat \Sigma v_{2} = \hat f \mbox{ in } \Omega \setminus \partial \Omega_3,
\end{equation}
and, on $\partial \Omega_3$,
\begin{align*}
\hat A \nabla v_2 \cdot \eta \Big|_\mathrm{ext} - \hat A \nabla v_2 \cdot \eta \Big|_\mathrm{int} &= A \nabla v \cdot \eta \Big|_\mathrm{ext} +  F_* A \nabla v_1 \cdot \eta \Big|_\mathrm{ext}  \quad ( \mbox{by } \eqref{reflextion}) \\[6pt]
& = A \nabla v \cdot \eta \Big|_\mathrm{ext} - F_* A \nabla v_1 \cdot \eta \Big|_\mathrm{int} \quad (\mbox{by } \eqref{eq-v1})\\[6pt]
& = A \nabla v \cdot \eta \Big|_\mathrm{ext} - F_* A \nabla (v + {\bf U}) \cdot \eta \Big|_\mathrm{int} \quad ( \mbox{by } \eqref{def-bfU}). 
\end{align*}
Since $F_{*} A = A$ in $\Omega_{3} \setminus \bar \Omega_{2}$, it follows from \eqref{bfU-U} that
\begin{equation}\label{eq-v2-bd1}
\hat A \nabla v_2 \cdot \eta \Big|_\mathrm{ext} - \hat A \nabla v_2 \cdot \eta \Big|_\mathrm{int}
 = - A \nabla  U \cdot \eta \Big|_\mathrm{int} \mbox{ on } \partial \Omega_{3}.
\end{equation}
Since $G(x) = x$ on $\partial \Omega_{3}$, we  obtain, on $\partial \Omega_3$,
\begin{equation}\label{eq-v2-bd2}
v_2\Big|_\mathrm{ext} - v_2\Big|_\mathrm{int} =  v\Big|_\mathrm{ext} - v_1\Big|_\mathrm{ext}  = v\Big|_\mathrm{ext} - v_1\Big|_\mathrm{int}  = v\Big|_\mathrm{ext} - ({\bf U} + v)\Big|_\mathrm{int}  = - U,  
\end{equation}
by \eqref{bfU-U}. Combining \eqref{eq-v2}, \eqref{eq-v2-bd1}, \eqref{eq-v2-bd2}, and \eqref{def-W-fre}, and applying the unique continuation principle, we have
\begin{equation}\label{v2-W}
v_2 = W \mbox{ in } \Omega.
\end{equation}
Since $\dive(A \nabla v) + k^{2} \Sigma v = f$ in  $\Omega_3 \setminus \bar \Omega_2$, it follows from  \eqref{def-v2} and \eqref{v2-W} that  
\begin{equation*}\left\{
\begin{array}{lc}
\dive(A \nabla v) + k^{2} \Sigma v= f & \mbox{ in } \Omega_3 \setminus \bar \Omega_2, \\[6pt]
v= W \Big|_{\mathrm{ext}} & \mbox{ on } \partial \Omega_3, \\[6pt]
A \nabla v  \cdot \eta \Big|_\mathrm{int} = A \nabla W \cdot \eta \Big|_{\mathrm{ext}} & \mbox{ on } \partial \Omega_3.
\end{array}\right.
\end{equation*}
By the unique continuation principle, it follows  from  \eqref{def-V-fre} that
\begin{equation}\label{v-V}
v = V \mbox{ in } \Omega_3 \setminus \bar \Omega_2.
\end{equation}
We claim that
\begin{equation*} v =\left\{
\begin{array}{cl}
W & \mbox{ in } \Omega \setminus \Omega_3, \\[6pt]
V &\mbox{ in }  \Omega_3 \setminus \Omega_2, \\[6pt]
(V+U) \circ F & \mbox{ in } \Omega_2 \setminus \Omega_1, \\[6pt]
W \circ G \circ F & \mbox{ in } \Omega_1. 
\end{array}\right.
\end{equation*}
In fact, the statement $v = W$ in $\Omega \setminus \Omega_{3}$ is a consequence of \eqref{def-v2} and \eqref{v2-W}; the statement
$v = V$ in $\Omega_{3} \setminus \Omega_{2}$ follows from \eqref{v-V}; the statement $v = (V+U) \circ F$ in $\Omega_{2} \setminus \Omega_{1}$ is a consequence of the fact $v_{1} = v + U = V + U$ in $\Omega_{3} \setminus \Omega_{2}$, and the statement $v = v_{1} \circ F $ in $\Omega_{2} \setminus \Omega_{1}$; $v =W \circ G \circ F $ in $\Omega_{1}$ is a consequence of the definition of $v_{1}$ and $v_{2}$, and $v_{2} = W$ in $\Omega_{3}$.  
The claim is proved. Therefore,
\begin{equation*}
v = NI(f) \mbox{ in } \Omega.
\end{equation*}
The proof of Step 1 is complete.

\medskip
 \underline{Step 2:}  We claim that
\begin{equation}\label{verify-eq-0}
\dive (A \nabla u_0) + k^{2} \Sigma u_{0}= f \mbox{ in } \Omega  \setminus (\partial \Omega_3 \cup \partial \Omega_2 \cup \partial \Omega_{1}).
\end{equation}
where $u_0 : = NI(f)$. Indeed, it is just a consequence of the definition of $U, \, V$ and $W$ and the fact that $F_{*} A = A$ and $F_{*} \Sigma = \Sigma$ in $\Omega_{3} \setminus \Omega_{2}$.

It remains to verify
\begin{equation}\label{cond-O3-O2}
[A \nabla u_0] = [u_0] = 0 \mbox{ on } \partial \Omega_3, \quad [\epss_0 A \nabla u_0] = [u_0] = 0 \mbox{ on } \partial \Omega_2, 
\end{equation}
and
\begin{equation}\label{cond-O1}
 [A \nabla u_0] = [u_0] = 0 \mbox{ on } \partial \Omega_1. 
\end{equation}
From the definitions of $V$ in  \eqref{def-V-fre} and $NI(f)$  in \eqref{def-NI}, we have
\begin{equation*}
[A \nabla u_0] = [u_0] = 0 \mbox{ on } \partial \Omega_3.
\end{equation*}
Since $U = 0$ and $A \nabla U \cdot \eta = 0$ on $\partial \Omega_2$, it follows from \eqref{reflextion} that
\begin{equation*}
 [\epss_0 A \nabla u_0] = [u_0] = 0 \mbox{ on } \partial \Omega_2. 
\end{equation*}
From \eqref{def-W-fre} and  \eqref{reflextion}, we have
\begin{equation*}
 [\epss_0 A \nabla u_0] = [u_0] = 0 \mbox{ on } \partial \Omega_1. 
\end{equation*}

The proof of Step 2 is complete. 

\medskip
\underline{Step 3:} Set
\begin{equation*}
v_\delta = u_\delta - u_0 \mbox{ in } \Omega.
\end{equation*}
We have, in $\Omega$,
\begin{align*}
\dive (\epss_\delta A \nabla v_\delta) + k^{2} \epss_{0} \Sigma v_{\delta} = & \dive (\epss_\delta A \nabla u_\delta) - \dive (\epss_\delta A \nabla u_0) + k^{2} \epss_{0} \Sigma u_{\delta} - k^{2} \epss_{0} \Sigma u_{0}
\end{align*}
This implies 
\begin{equation*}
\dive (\epss_\delta A \nabla v_\delta) + k^{2} \epss_{0} \Sigma v_{\delta} =  \dive \big[(\epss_0 - \epss_\delta) A \nabla u_0\big] \mbox{ in } \Omega.
\end{equation*}
By Lemma~\ref{lem1}, we have
\begin{equation*}
\| \nabla v_\delta \|_{L^2(\Omega)} \le C \|\nabla u_0 \|_{L^2(\Omega)},
\end{equation*}
which yields, since $u_\delta = v_\delta + u_0$,
\begin{equation*}
\| \nabla u_\delta \|_{L^2(\Omega)} \le C \|\nabla u_0 \|_{L^2(\Omega)}.
\end{equation*}
Since $u_\delta \in H^1_0(\Omega)$, by Poincar\'e's inequality, it follows that
\begin{equation*}
\| u_\delta \|_{H^1(\Omega)} \le C \|\nabla u_0 \|_{L^2(\Omega)}.
\end{equation*}
Step 3 completes.

\medskip
\underline{Step 4:} The conclusion of Step 4 follows from Step 3 and the fact that the limit of $u_\delta$ (up to a subsequence) satisfies \eqref{eq-lim0} and \eqref{eq-lim0} has a unique solution in $H^1_0(\Omega)$ by Steps 1 and 2.  \proofend

\subsubsection{Proof of the second statement of Theorem~\ref{thm2}}

In this section $f$ is not compatible. We prove the second statement of Theorem~\ref{thm2} by contradiction. Assume that \eqref{limit-energy-k} is not true. Without loss of generality, there exists  a bounded sequence $(u_\delta)$ in $H_{0}^1(\Omega)$  such that $u_{\delta}$ is the unique solution to the equation 
\begin{equation*}
\dive(\epss_{\delta} A \nabla u_{\delta}) + k^{2} \epss_{0} \Sigma u_{\delta} = \epss_{0} f \mbox{ in } \Omega, 
\end{equation*}
and 
$u_\delta$ converges weakly to (some) $u \in H^1(\Omega)$ as $\delta \to 0$. It follows that $u \in H^1_0(\Omega)$ is a solution to the equation
\begin{equation*}
\dive(\epss_0 A \nabla u) + k^{2} \epss_{0} u= \epss_0 f \mbox{ in } \Omega.
\end{equation*}
Define
\begin{equation*}
U= u \circ F^{-1} - u \mbox{ in } \Omega_3 \setminus \bar \Omega_2 \quad \mbox{ and } \quad V = u \mbox{ in } \Omega_3 \setminus \bar \Omega_2.
\end{equation*}
As in Step 1 of Section~\ref{sec-thm2}, $U$ and $V$ satisfy \eqref{def-U-fre} and \eqref{def-V-fre} respectively. We have a contradiction since $f$ is not compatible with the system. \proofend

\bigskip
\noindent {\bf Acknowledgment.} The author would like to thank Bob Kohn and Grame Milton for interesting discussions.



\providecommand{\bysame}{\leavevmode\hbox to3em{\hrulefill}\thinspace}
\providecommand{\MR}{\relax\ifhmode\unskip\space\fi MR }
\providecommand{\MRhref}[2]{%
  \href{http://www.ams.org/mathscinet-getitem?mr=#1}{#2}
}
\providecommand{\href}[2]{#2}

\end{document}